# Modified One-Axis-Twist Squeezing for Deterministic Orientation Control of Schrödinger Cat States with Unknown Atom-Number Parity

Jinyang Li[1] and Selim M Shahriar[1,2]

[1]Department of Physics and Astronomy, Northwestern University, Evanston, IL 60208, USA
[2]Department of Electrical and Computer Engineering, Northwestern University, Evanston, IL 60208, USA

**Abstract**

The Schrödinger cat state protocol (SCSP) makes use of a Schrödinger cat (SC) state generated with one-axis-twist squeezing (OATS) to enhance the sensitivity. In principle, the SCSP can magnify the phase shift by a factor *N*, the number of atoms under interrogation, accompanied by a quantum noise amplification by a factor of root-*N*, enabling an atomic sensor to reach the Heisenberg limit. However, the current SCSP can reach this benchmark only if the parity of *N* is known, which is almost impossible for a large ensemble. The reason is the orientation of the SC state generated with the conventional OATS depends on the parity of *N*. In this paper, we propose a modified OATS (MOATS) operation that generates the SC state aligned in a certain direction regardless of the parity of *N*.

## 1. Introduction

Spin squeezing [1] is a promising technique for pushing quantum sensors towards the Heisenberg limit. One-axis-twist squeezing (OATS) is the most experimentally accessible type of spin squeezing. For concreteness, this paper is restricted to only OATS. The OATS is described by the Hamiltonian $\hbar\chi S_z^2$ and the corresponding propagator $e^{-i\mu S_z^2}$, where $S_z$ is the *z* component of the dimensionless spin operator representing the sum of the spin operators of all the atoms, $\chi$ is the strength of the nonlinear interaction, and $\mu$ equals $\chi$ times the interaction time. The resulting state after an OATS operation is applied to a coherent spin state (CSS) oriented in the *x* direction is as follows. As $\mu$ increases from zero, the uncertainty of the spin operator in a certain direction decreases at the expense of an increase in the uncertainty along a perpendicular direction. As $\mu$ continues to increase, the resulting state becomes more complicated. For special cases when $\mu = \pi/p$ ($p \in \mathbb{Z}_+$), the resulting state is a superposition of $p$ CSSs evenly distributed around the equator of the Bloch sphere [2]. More specially, for $\mu = \pi/2$, the resulting state is a Schrödinger cat (SC) state consisting of two CSSs oriented in the opposite direction. OATS can be used for magnifying quantum phase shift (i.e., narrowing the signal fringes as a function of the phase shift) if it is applied in the echo configuration. The protocols in such a configuration are named echo squeezing protocols (ESPs).

The ESPs in principle can be adapted to any atomic sensor employing coherent quantum states, for example, Ramsey clocks [3, 4], Ramsey magnetometers [5], and atom interferometers [6, 7]. There are three types of ESPs, which produce different degrees of phase magnifications and noise amplifications. The first one is the conventional ESP (CESP) [8], which can ideally magnify the phase shift by $\sqrt{N/e}$ ($N$ being the number of the atoms under interrogation), while leaving the quantum noise unchanged, for a particular value of $\mu$. The second one is the generalized ESP (GESP) [3], which can ideally magnify the phase shift by $N \sin \mu/\sqrt{2}$, accompanied by an amplification in the quantum noise level by $\sqrt{N} \sin \mu$, for a broad range of values of $\mu$. The third one is the SC state protocol (SCSP) [3,4], which can ideally magnify the phase shift by the factor of $N$, while amplifying the quantum noise level by the factor of $\sqrt{N}$, for $\mu = \pi / 2$, if the parity (i.e., whether $N$ is even or odd) is known a priori.

The original SCSP proposed in Ref. [4] is optimized for only even parity of $N$, and Ref. [3] shows two versions of the SCSP optimized for even and odd values of $N$. If the parity of $N$ matches the protocol designed for it, the phase shift is magnified by $N$, and the noise level is amplified by $\sqrt{N}$, enabling the sensor to reach the Heisenberg limit. Otherwise, both the phase magnification factor and the noise amplification factor equal $\sqrt{N}$, and the sensitivity remains at the standard quantum limit. The reason for the dependence on the parity of $N$ is that the orientation of the Schrödinger cat state is determined by the parity of $N$ for the conventional OATS (COATS). Starting with a CSS oriented in the $x$ direction, for example, the final state is an $x$ directed SC state if $N$ is even, and is a $y$ directed SC state if $N$ is odd. For a large number of atoms in an experiment, it is essentially impossible to know the exact value of $N$. Although the signal does not disappear even if parity-matched and parity-mismatched shots are mixed, the parity-mismatched shots represents a waste of resources and measurement time. In this paper, we propose a parity-independent SCSP. Specifically, we show that when the OATS operation is modified in a particular way, the orientation of the SC state becomes independent of the parity of $N$.

## 2. Modified OATS

For specificity, and without loss of generality, we assume that the atoms are initially prepared in the CSS oriented in the $x$ direction. A COATS operator for $\mu = \pi/2$ generates an $x$ directed SC state [2, 9], which can be mathematically expressed as

$$e^{-i(\pi/2)S_z^2} \left|\mathbf{x}\right\rangle = e^{i\alpha_0} \left( \left|\mathbf{x}\right\rangle + e^{i\phi_0} \left|-\mathbf{x}\right\rangle \right) \tag{1}$$

where $\left|\mathbf{x}\right\rangle$ represents the CSS oriented in the $x$ direction, $\left|-\mathbf{x}\right\rangle$ represents the CSS oriented in the $-x$ direction, and the exact expressions for the phase factors $\alpha_0$ and $\phi_0$ can be found in Ref. [2]. A COATS operator for $\mu = \pi/2$ generates an $y$ directed SC state, which can be mathematically expressed as

$$e^{-i(\pi/2)S_z^2} \left|\mathbf{x}\right\rangle = e^{i\alpha_1} \left( \left|\mathbf{y}\right\rangle + e^{i\phi_1} \left|-\mathbf{y}\right\rangle \right) \tag{2}$$

where $\left|\mathbf{y}\right\rangle$ represents the CSS oriented in the $y$ direction, $\left|-\mathbf{y}\right\rangle$ represents the CSS oriented in the $-y$ direction, and the exact expressions for the phase factors $\alpha_1$ and $\phi_1$ can be found in Ref. [2].

The simplest version of the modified OATS (MOATS) is described by the Hamiltonian $\hbar\chi N_\uparrow^2$ and the corresponding propagator $e^{-i\mu N_\uparrow^2}$, where $N_{\uparrow(\downarrow)}$ is the operator describing the number of atoms in the $\left|\uparrow(\downarrow)\right\rangle$ state, which are related to $S_z$ according to the expression $S_z = \left(N_\uparrow - N_\downarrow\right)/2$. For even values of $N$, the state after the MOATS operation is as follows:

$$
\begin{aligned}
e^{-i(\pi/2)N_\uparrow^2}\left|\mathbf{x}\right\rangle &= \exp\left[-i\frac{\pi}{2}\left(\frac{N}{2}+S_z\right)^2\right]\left|\mathbf{x}\right\rangle = e^{-i\pi N^2/8}e^{-i(\pi N/2)S_z}e^{-i(\pi/2)S_z^2}\left|\mathbf{x}\right\rangle \\
&= e^{i(\alpha_0-\pi/8)}e^{-i(\pi N/2)S_z}\left(\left|\mathbf{x}\right\rangle + e^{i\phi_0}\left|-\mathbf{x}\right\rangle\right) \\
&= e^{i(\alpha_0-3\pi/8)}\begin{cases}\left|\mathbf{x}\right\rangle + e^{i\phi_0}\left|-\mathbf{x}\right\rangle, & (N/2 \text{ is even}) \\ e^{i\phi_0}\left|\mathbf{x}\right\rangle + \left|-\mathbf{x}\right\rangle, & (N/2 \text{ is odd})\end{cases}
\end{aligned} \tag{3}
$$

The last step makes use of the fact that that $e^{-i(\pi N/2)S_z}$ is a rotation operator that rotates a quantum state by $\pi N/2$ and that $N$ is even. The final state after this MOATS operation is still an $x$ directed SC state.

It should be noted that the final state after the MOATS operation depends on the parity of $N/2$ in a trivial way. Namely, the relative phase difference between the two components of the SC state differs by $2\phi_0$ depending on the parity of $N/2$. However, in either case, the final state is still an x directed SC state, and this phase difference does not affect the sensitivity of a sensor employing the MOATS protocol.

For odd values of $N$, the state after the MOATS operation is derived below.

$$
\begin{aligned}
e^{-i(\pi/2)N_\uparrow^2}\left|\mathbf{x}\right\rangle &= \exp\left[-i\frac{\pi}{2}\left(\frac{N}{2}+S_z\right)^2\right]\left|\mathbf{x}\right\rangle = e^{-i\pi N^2/8}e^{-i(\pi N/2)S_z}e^{-i(\pi/2)S_z^2}\left|\mathbf{x}\right\rangle \\
&= e^{i(\alpha_1-\pi/8)}e^{-i(\pi N/2)S_z}\left(\left|\mathbf{y}\right\rangle + e^{i\phi_1}\left|-\mathbf{y}\right\rangle\right) \\
&= e^{i(\alpha_1-3\pi/8)}\begin{cases}\left|\mathbf{x}\right\rangle + e^{i\phi_1}\left|-\mathbf{x}\right\rangle, & \left(\dfrac{N+1}{2} \text{ is even}\right) \\ e^{i\phi_1}\left|\mathbf{x}\right\rangle + \left|-\mathbf{x}\right\rangle, & \left(\dfrac{N+1}{2} \text{ is odd}\right)\end{cases}
\end{aligned} \tag{4}
$$

Here again it should be noted that the final state after the MOATS operation depends on the parity of $(N+1)/2$ in a trivial way. Namely, the relative phase difference between the two components of the SC state differs by $2\phi_1$ depending on the parity of $(N+1)/2$. However, in either case, the final state is still an x directed SC state, and this phase difference does not affect the sensitivity of a sensor employing the MOATS protocol.

This equation shows the key difference between the MOATS and the COATS. Even if $N$ is odd, the final state after the MOATS operation is still an $x$ directed SC state. Therefore, the SCSP optimized for even values of $N$ described in Ref. [3, 4] is now optimized for both parities if the MOATS takes the place of the COATS.

Following the same steps, it can be proven that the MOATS described by the Hamiltonian $\hbar\chi\left(2N_{\uparrow}+N_{\downarrow}\right)^2$ can also generate parity-independent SC state. This version of the MOATS is also important because it can be implemented with cavity-mediated nonlinear interaction, while the simplest version cannot. The practical methods for implementing MOATS is discussed in the next section.

## 3. Practical methods for implementing the MOATS

Currently, there are two primary methods for implementing OATS, namely cavity-mediated [10, 11, 12, 13, 14] and Rydberg-mediated [15, 16, 17] nonlinear interactions. It turns out that the Rydberg-mediated nonlinear interaction is intrinsically the simplest version of the MOATS, while cavity-mediated nonlinear interaction can only be used for more complicated versions of the MOATS.

For Rydberg-mediated nonlinear interaction, an atom is modeled as a three-level system consisting of two ground states (corresponding to the $|\uparrow\rangle$ and the $|\downarrow\rangle$ state), and a Rydberg state $|r\rangle$. Only the $|\uparrow\rangle$ state is coupled to the Rydberg state with a far detuned optical field. The detuning is denoted by $\Delta$ and the Rabi frequency by $\Omega$. Due to Rydberg blockade, only one atom can be excited to the Rydberg state, so that any collective state with more than one atom in the $|r\rangle$ state will not be populated. Consider the energy of a Rydberg dressed state with $N_{\uparrow}$ atoms in the $|\uparrow\rangle$ state and $N_{\downarrow}$ atoms in the $|\downarrow\rangle$ state. The coupling strength between the bare collective state and the collective state with one atom in the $|r\rangle$ state is given by $\sqrt{N_{\uparrow}}\Omega$ [18]. The corrections to the energy can be calculated with perturbation theory [19]. Odd orders of corrections to the energy are found to be zero. The second order correction is given by $\hbar N_{\uparrow}\Omega^2/4\Delta$, which is obviously the light shift. The fourth order correction is given by $-\hbar N_{\uparrow}^2\Omega^4/16\Delta^3$, which is exactly the Hamiltonian for the MOATS, with $\chi=-\Omega^4/16\Delta^3$.

To implement MOATS with cavity-mediated nonlinear interaction, the squeezing beam is tuned to such a frequency that the light shift of the $|\uparrow\rangle$ state is twice that of the $|\downarrow\rangle$ state. It is shown in Appendix B of Ref. [5] that such a system can be described by the non-Hermitian Hamiltonian

$$H = \hbar\left[-\delta + \varepsilon\left(2N_{\uparrow} + N_{\downarrow}\right) - i\frac{\kappa}{2}\right]a^{\dagger}a + i\hbar\sqrt{\kappa_0}\left(a^{\dagger}\beta - a\beta^{*}\right) \tag{5}$$

where $\delta$ is the detuning of the probe from the cavity resonant frequency, $\varepsilon$ is the single-photon-induced light shift of the $|\downarrow\rangle$ state, $a$ is the annihilation operator for the cavity field, $\beta$ is defined in such a way that $|\beta|^{2}$ equals the number of incident photons per unit time, the complex argument of $\beta$ equals the phase of the incident beam, $\kappa$ is the total decay rate of the cavity, and $\kappa_0$ is the decay rate resulting from the transmission of the input mirror. Eliminating the optical field according to the steps illustrated in Appendix B of Ref. [5] yields the effective Hamiltonian $i\hbar\sqrt{\kappa_0}\beta^{*}\left(\alpha - \alpha'\right)$, where

$$\alpha = \frac{i\sqrt{\kappa_0}\beta}{\delta - i\kappa/2}; \quad \alpha' = \frac{i\sqrt{\kappa_0}\beta}{\delta - \varepsilon\left(2N_\uparrow + N_\downarrow\right) + i\kappa/2} \tag{6}$$

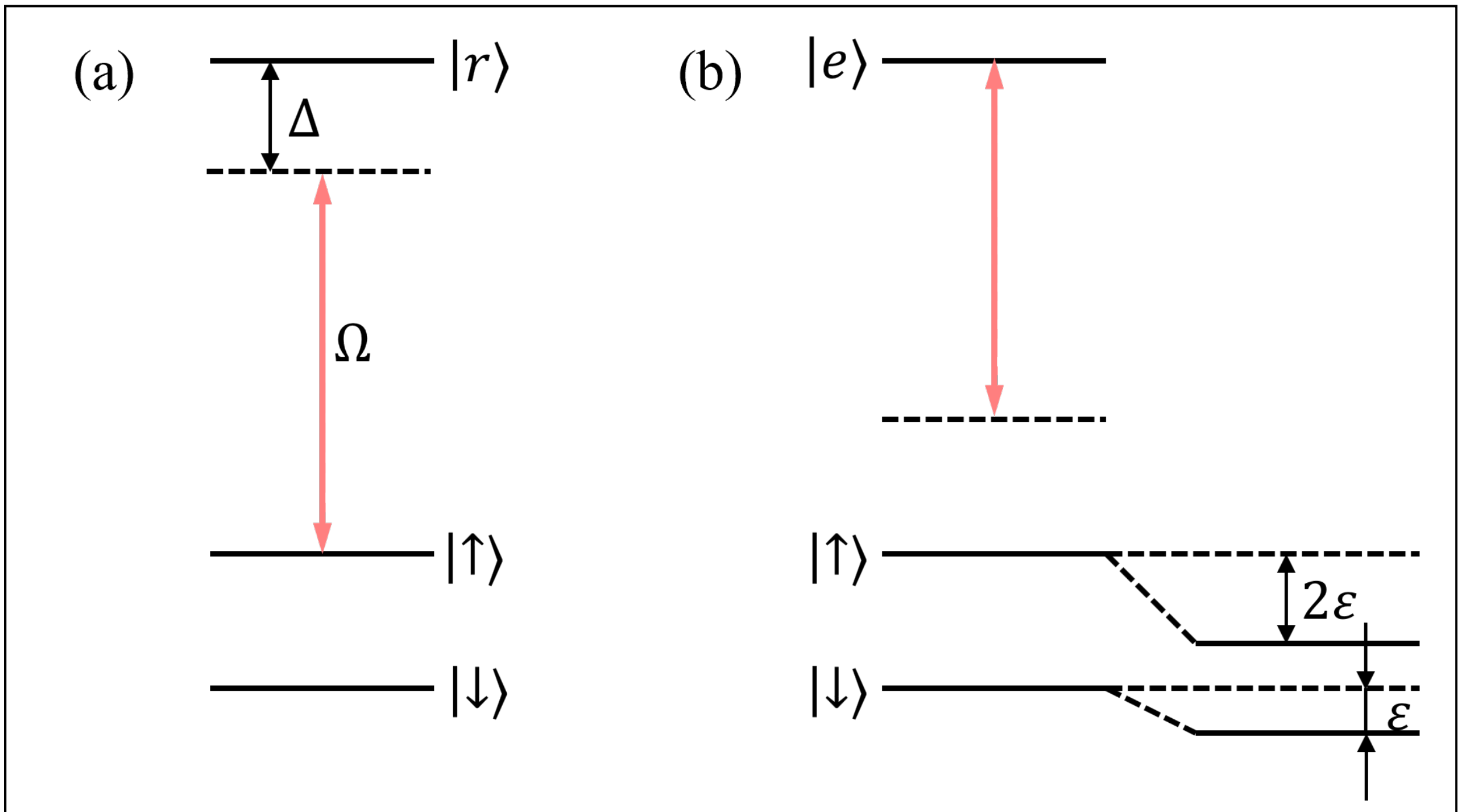

Figure 1. (a) Scheme for realizing the version of the MOATS described by the Hamiltonian $\hbar\chi N_{\uparrow}^{2}$ with Rydberg-mediated nonlinear interaction. The optical field coupling the $|\uparrow\rangle$ and the $|r\rangle$ state is far detuned. The detuning is denoted by $\Delta$ and the Rabi frequency by $\Omega$. (b) Scheme for realizing the version of the MOATS described by the Hamiltonian $\hbar\chi(2N_{\uparrow} + N_{\downarrow})^{2}$ with cavity-mediated nonlinear interaction.

Keeping the linear and the quadratic term of $\varepsilon$ yields

$$H = \hbar\varepsilon|\alpha|^2\left(2N_\uparrow + N_\downarrow\right) + \frac{\hbar\delta\varepsilon^2|\alpha|^2}{\delta^2+\kappa^2/4}\left(2N_\uparrow + N_\downarrow\right)^2 - i\hbar\frac{\kappa}{2}\frac{\varepsilon^2|\alpha|^2}{\left(\delta^2+\kappa^2/4\right)}\left(2N_\uparrow + N_\downarrow\right)^2 \tag{7}$$

The linear term describes the light shift induced by the unperturbed cavity field, the real part of the quadratic term is exactly the Hamiltonian for the MOATS, with $\chi = \frac{\delta\varepsilon^2|\alpha|^2}{\delta^2+\kappa^2/4}$, and the imaginary part describes the decoherence mechanism due to cavity decay.

## 4. Conclusion

Previously, we had shown that the Schrödinger cat state protocol (SCSP) makes use of a Schrödinger cat (SC) state generated with one-axis-twist squeezing (OATS) to enhance the sensitivity. In principle, the SCSP can magnify the phase shift by a factor $N$, the number of atoms under interrogation, accompanied by a quantum noise amplification by a factor of $\sqrt{N}$, enabling an atomic sensor to reach the Heisenberg limit. However, the current SCSP can reach this benchmark only if the parity of $N$ is known, which is virtually impossible for a large ensemble. The reason is the orientation of the SC state generated with the conventional OATS (COATS) depends on the parity of $N$. In this paper, we have proposed a modified OATS (MOATS) operation that generates the SC state aligned in a certain direction regardless of the parity of $N$. We present two different squeezing Hamiltonians for the MOATS. The simpler version of the MOATS can be implemented using Rydberg blockade induced non-linear interaction, while the more complicated version of the MOATS can be implemented with the cavity mediated non-linear interaction. Additional analysis of these two MOATSs are needed to determine how the overall sensitivity achievable using these compare with what can be achieved using the COATS when effects of various sources of noise induced by the squeezing interaction are taken into account. This work is currently underway, and will be reported in the near future.

**Acknowledgement:**

This work has been supported equally in parts by the Department of War Center of Excellence in Advanced Quantum Sensing under Army Research Office grant number W911NF202076, ONR grant number N00014-19-1-2181, and NASA NIAC grant number 80NSSC25K8008.

[18] Saffman, M., Walker, T. G., & Mølmer, K. (2010). Quantum information with Rydberg atoms. Reviews of modern physics, 82(3), 2313-2363.

[19] Li, J. (2024). Large Momentum Transfer and Spin Squeezing for Atom Interferometry (Doctoral dissertation, Northwestern University), pp. 21-23